\documentclass[12pt]{article}

\usepackage{amsmath,amsfonts}

\usepackage{slashed}

\advance \textheight by \topskip
\makeatletter
\@addtoreset{equation}{section}
 
\makeatother

\newcommand{\be}{\begin{equation}}
\newcommand{\ee}{\end{equation}}
\newcommand{\bea}{\begin{eqnarray}}
\newcommand{\eea}{\end{eqnarray}}
\newcommand{\dd}{\partial}

\def\>{\rangle}
\def\<{\langle}

\begin{document}

\title{{\bf On Higgs inflation in non-minimally coupled models of gravity }}

\author{
{\sf   N. Mohammedi} \thanks{e-mail:
noureddine.mohammedi@univ-tours.fr}$\,\,$${}$
\\
{\small ${}${\it Institut  Denis Poisson (CNRS - UMR 7013),}} \\
{\small {\it Universit\'e  de Tours,}}\\
{\small {\it Facult\'e des Sciences et Techniques,}}\\
{\small {\it Parc de Grandmont, F-37200 Tours, France.}}}

\date{}
\maketitle
\vskip-1.5cm

\vspace{2truecm}

\begin{abstract}

\noindent
Models of inflation in which the Higgs field is non-minimally coupled to gravity lead, after a recaling of the metric,  
to a complicated expression for the potential of the inflaton field. Nevertheless, this potential produces the desired features, both
theoretical and  experimental,
after some approximations are made. In this note, we also allow for a  modification of  the Higgs kinetic term in such a way that the resulting  potential
for the inflaton field 
is, without any approximation,  of the simplest form after the rescaling of the metric.

\end{abstract}

\newpage

\setcounter{equation}{0}

\section{Introduction}

The Standard Model of particles physics has proved to be a robust theory in describing
the observed interactions between particles at the level of particle accelerators. However, 
this successful theory does not seem to answer questions related to the cosmological evolution of the 
Universe. In particular, the nature of dark matter or dark energy as well as the
early inflationary regime of the Universe remain a mystery. 
In this note we will be mostly concerned  with the problem of inflation. 
Introductory reviews of cosmic inflation can be found in \cite{liddle, tsujikawa,senatore,mughal,julien}.
\par
It is widely accepted now that
inflation might be driven by a scalar field moving slowly in a certain potential.
The discovery of the Higgs field has completed the Standard Model and opened a new 
window in cosmology. It is therefore tempting to identify the Higgs field  with the scalar field needed for inflation.
\par
The issue now is how to couple the Standard Model to gravity in order to account for inflation
through the Higgs field.
Various models can be constructed depending on what kind of coupling one chooses. Accounts of the Higgs field
in cosmology can be found in \cite{steinwachs,rubio,horn}. Here we will mention only what is called the non-minimal Higgs inflation model as 
it is relevant to our study. In this theory, the coupling of the Higgs field to gravity is given 
by the action \cite{kaiser, bezrukov1,bezrukov2}
\bea
S = 
\int d^4x\sqrt{- {g}}\left[- \frac{1}{2}\left(
M^2 +{2\xi}\,H^\dagger H\right)R +{\cal L}_{\text{SM}} \right]
 \,\,\,.
\eea
The quantity ${\cal L}_{\text{SM}}$ is the full Standard Model Lagrangian written in curved space-time
and $H$ is the Higgs doublet. We will return to this model of inflation in some details in section 3.
It suffices to say for the moment that under a certain approximation (see later), this theory gives a viable
explanation of inflation. 
\par
In order to get the  standard term $\int d^4x\sqrt{- {g}} R $ for the gravitational sector,  the authors of refs.\cite{kaiser, bezrukov1,bezrukov2}
carried out a conformal rescaling of the metric field. This operation induced a complicated expression 
for the potential of the Higgs field (or equivalently, gave a non-canonical kinetic term for the Higgs field).  
Nevertheless, this complication disappears under the above cited approximation \cite{bezrukov1,bezrukov2} .
\par
The aim of this note is to provide a model of Higgs inflation where no approximation is needed.
We have simply modified the Higgs kinetic term in the Standard Model non-minimally coupled to gravity.
More specifically, we propose the model described by the action
\bea
S = 
\int d^4x\sqrt{- {g}}\left[- \frac{1}{2}\left(
M^2 +{2\xi}\,H^\dagger H\right)R +
F\left(H^\dagger H\right)\left(D_\mu H\right)^\dagger\left(D^\mu H\right)+
{\cal L}_{\text{SM}}^\prime \right]
 \,\,\,
\eea
to account for  inflation of the early Universe.
Here ${\cal L}_{\text{SM}}^\prime$ is the Standard Model Lagrangian without the gauge covariant kinetic term
and the function $F\left(H^\dagger H\right)$ is determined in such a way that the resulting
kinetic term of the inflaton field, after a rescaling of the metric, is in  the usual canonical form. 
This theory could be seen as an improvement of the proposition of ref.{\cite{bezrukov1}}. It is also crucial 
to mention that the two models coincide for a large Higgs field (which is precisely the 
approximation made in ref.{\cite{bezrukov1}}). 
\par
The article is organised as follows: In the next section we review the Starobinsky model of inflation
\cite{starobinsky}
as there are similarities between this model and non-minimal Higgs inflation \cite{kehagias}.
Some details of the non-minimal coupling of the Higgs field to gravity are given in section 3.
Our modification of the non-minimally coupled Higgs field is presented in section 4. 
The last section is dedicated to some conclusions.

\section{The Starobinsky model of inflation : An inspiration}

Various models dealing with the issue of the rapid expansion of the early Univers have been 
inspired by the  Starobinsky model of inflation \cite{kehagias}. This is described by the action \cite{starobinsky}
\bea
S_0= 
\int d^4x\sqrt{-{g}}\left[-\frac{M^2_\text{p}}{2}\, {R} + \frac{1}{12M^2_{\text S}} \, {R}^2 \right]
 \,\,\,.
\label{Starobinsky}
\eea
As it is well-known, the quadratic term in the scalar curvature can be traded for a scalar field \cite{whitt}. 
One starts by writing the equivalent action
\bea
S_1= 
\int d^4x\sqrt{-{g}}\left[-\frac{M^2_\text{p}}{2} \left(1  
+ \frac{\varphi}{3 M^2_\text{p} M^2_{\text S} } \right){R} - \frac{\varphi^2 }{12 M^2_{\text S}} \right]
 \,\,\,,
\label{Starobinsky-1}
\eea
where $\varphi$ is an auxiliary field whose
equation of motion is $\varphi=- R$. Upon 
replacing this  in the last action we recover the Starobinsky model (\ref{Starobinsky}).
\par
The action (\ref{Starobinsky-1}) can be cast in a more familiar form by going to the Einstein frame
through the field redefinition
\bea
 g_{\mu\nu} \longrightarrow e^{\omega X(x)}\, {g}_{\mu\nu} 
\,\,\,\,\,\,\,\,\,\,\,\,{\text{with}}\,\,\,\,\,\,\,\,\,\,\,\,
\left(1  + \frac{\varphi}{3 M^2_\text{p} M^2_{\text S} } \right) 
\,e^{\omega X(x)}=1 
\,\,\,\,\,\,,\,\,\,\,\,\,\,
\omega  =  -\, \sqrt{\frac{2}{3}}\,\frac{1}{M_\text{p} }
\,\,\,\,.
\eea
With this conformal rescaling of the metric, the action  (\ref{Starobinsky-1}) yields (up to a total
derivative)
\bea
S_2= 
\int d^4x\sqrt{- {g}}\left[-\frac{M^2_\text{p}}{2}\,{R} 
+ \frac{1 }{2}\, {g}^{\mu\nu}\dd_\mu X\,\dd_\nu X
-\frac{3}{4}\,M^4_\text{p}M^2_{\text S} \left(1- \,e^{\omega X}\right)^2
 \right]
 \,\,\,.
\label{Starobinsky-2}
\eea
The two actions (\ref{Starobinsky})  and (\ref{Starobinsky-2}) could be thought of as dual to 
each other.
\par
We see that the field  $X(x)$ is moving in a potential which is almost flat for very large values of  $X(x)$. 
This is the most sought feature which realises the slow-roll condition needed for inflationary models. In this case
the kinetic term of the field  $X(x)$ is negligeable and the cosmological evolution of the Universe is dominated 
by the constant term  $\frac{3}{4}\,M^4_\text{p}M^2_{\text S}$ (playing the role of a cosmological constant). 
The Starobinsky model is compatible with cosmological measurements
for $M_{\text S} \approx 10^{-5}$.
\par
It is essentially the theory in (\ref{Starobinsky-2}) which was reached by the non-minimally coupled Higgs field of ref.\cite{bezrukov1}.
It was also shown in \cite{kehagias} that other models of inflation could be thought of, under certain assumptions,  as descendants 
of the Starobinsky model.

\section{The Higgs field non-minimally coupled to gravity}

Here and in the rest of this note,  we will assume that the Higgs doublet is in the unitary gauge
\be
H\left(x\right)= \frac{1}{\sqrt{2}}\left(
\begin{array}{l}
0\\  h\left(x\right) \end{array}\right)
\,\,\,.
\ee
The Higgs field non-minimally coupled to gravity is
described, in the Jordan frame,  by the action \cite{kaiser, bezrukov1,bezrukov2}
\bea
S_\text{J} = 
\int d^4x\sqrt{- {g}}\left[- \frac{1}{2}\left(
M^2 +{\xi}\,h^2\right)R +\frac{1}{2}\dd_\mu h\dd^\mu h
-\frac{\lambda}{4}\left(h^2-v^2\right)^2 \right]
 \,\,\,.
\label{B-S-0}
\eea
We have written down only the terms relevant to our study. The constant $M$ 
and the reduced  Planck mass ${M_\text{p} }=\frac{1}{\sqrt{8\pi G}}=2.4335\times 10^{18}$ GeV are related by 
\be
 {M_\text{p} }^2=M^2+\xi\,v^2
\ee
with the Higgs field expectation value $v=246$ GeV. We are assuming that $\hbar=c=1$.
\par
This theory, however,  is best studied in the Einstein frame by performing
the conformal transformation   
\bea
g_{\mu\nu} \longrightarrow e^{\omega\varphi} \,{g}_{\mu\nu}
\,\,\,\,\,\,\,\,\,\,\,\,{\text{with}}\,\,\,\,\,\,\,\,\,\,\,\,
\omega  =  -\, \sqrt{\frac{2}{{3}}}\,\frac{1}{M_\text{p} }
\,\,\,\,.
\eea
The fields $\varphi$ and ${h}$ are related by 
\bea
 \left(M^2+  {\xi}{{h}^2}  \right)e^{\omega\varphi} = {M^2_\text{p}}
\,\,\,\,.
\label{h-rela}
\eea
Therefore one can use either one  of the two fields to explore the cosmological
properties of the theory. We find it more convenient to keep the field  $\varphi$
instead of the field $h$ as it makes 
the various approximations  more transparent.

\par
In terms of the
field $\varphi$ and up to a total derivative,
the action in the Einstein frame is given by
\bea
S_\text{E} &=& 
\int d^4x\sqrt{- {g}}\bigg\{
- \frac{M^2_\text{p}}{2}\, {R} 
+\frac{3}{4}M^2_\text{p}\omega^2 \left[{1} + \frac{1}{6\xi}
\frac{1}{\left(1-\frac{M^2}{M^2_\text{p}}\, e^{\omega\varphi}\right) }
\right] {g}^{\mu\nu}\dd_\mu \varphi \dd_\nu \varphi
\nonumber\\
&-& \frac{\lambda}{4}\left(\frac{M^2_\text{p}}{\xi}\right)^2\bigg(1-  e^{\omega\varphi} \bigg)^2 
\bigg\}
 \,\,\,.
\label{B-S}
\eea
We have used  ${M_\text{p} }^2=M^2+\xi\,v^2$ in the last term.
We notice that the potentiel terms is readily in the much sought exponentiel form (as in the Starobinsky model).
This is precisely what one needs for inflation.
However, the kinetic term for the scalar field $\varphi$ is not standard. 
Nevertheless, in the case when 
\be
\frac{M^2}{M^2_\text{p}}\, e^{\omega\varphi} << 1
\,\,\,\,\,\,\,\,\,,\,\,\,\,\,\,\,\,\,
\xi>>1
\ee
the second term in the kinetic energy can be neglected and  one arrives
at the action\footnote{The approximation $\frac{M^2}{M^2_\text{p}}\, e^{\omega\varphi} << 1$ is the  slow-roll condition
and corresponds to $h^2>> \frac{M^2}{\xi}$ as can be seen from (\ref{h-rela}).}
\bea
S_\text{E} &\simeq& 
\int d^4x\sqrt{- {g}}\bigg\{
- \frac{M^2_\text{p}}{2}\, {R} 
+\frac{1}{2}
\, {g}^{\mu\nu}\dd_\mu \varphi \dd_\nu \varphi
- \frac{\lambda}{4}\left(\frac{M^2_\text{p}}{\xi}\right)^2\bigg(1-  e^{\omega\varphi} \bigg)^2 
\bigg\}
 \,\,\,.
\label{B-S-apprx}
\eea
This theory looks then very much like the Starobinsky model (\ref{Starobinsky-2}) 
with the identification $3M^2_{\text S}=\frac{\lambda}{\xi^2}$.
The observed 
cosmological data requires $\xi=1.8\times 10^4$ for a Higgs self-coupling\footnote{The constant $\lambda$ is related 
to the Higgs mass $m_{\text H}$ and the Higgs expectation value $v$ by the 
relation $\lambda=\frac{m^2_{\text H} }{2v^2}\simeq \frac{(126)^2 }{2\times (246)^2}\simeq 0.13$.}
$\lambda = 15\times 10^{-2}$ (see \cite {bezrukov1,bezrukov2}).
These values are also compatible with $M_{\text S} \approx 10^{-5}$.
\par
The authors of refs.\cite{kaiser, bezrukov1,bezrukov2} have adopted another route
in studying the model in (\ref{B-S-0}).
There, a standard kinetic term was obtained by introducing
a new field $\chi$ through
\bea
\left(\frac{d\chi}{d\varphi}\right)^2=  \frac{3}{2}M^2_\text{p}\omega^2 \left[{1} + \frac{1}{6\xi}
\frac{1}{\left(1-\frac{M^2}{M^2_\text{p}}\, e^{\omega\varphi}\right) }
\right]\,\,\,.
\label{chi-rel}
\eea
The action (\ref{B-S}) becomes then
\bea
S_\text{E} =  
\int d^4x\sqrt{- {g}}\left\{
- \frac{M^2_\text{p}}{2}\, {R} 
+\frac{1}{2} {g}^{\mu\nu}\dd_\mu \chi \dd_\nu \chi -U\left(\chi\right)\right\}
 \,\,\,.
\label{B-S-1}
\eea
The potential term $U\left(\chi\right)$ for the new field $\chi(x)$ is extracted from
\be
U\left(\chi\right)
=\frac{\lambda}{4}\left(\frac{M^2_\text{p}}{\xi}\right)^2\bigg(1-  e^{\omega\varphi} \bigg)^2 
\ee
by first integrating (\ref{chi-rel}) and then  expressing $\varphi$ in terms of $\chi$. 
This results in a quite involved expression for $U\left(\chi\right)$.
However, when $\frac{M^2}{M^2_\text{p}}\, e^{\omega\varphi} << 1$ 
and $\xi << 1$ one sees from (\ref{chi-rel}) that 
${\omega\varphi}\simeq  -\sqrt{\frac{2}{3}}\, \frac{1}{M_\text{p}}\,\chi$.
In this case the actions (\ref{B-S-1}) and  (\ref{B-S-apprx}) are the same
after the exchange $\varphi\leftrightarrow\chi$.

\section{A scalar-tensor gravity with the Higgs field}

In this section, we propose another modification of the coupling
of the Standard Model to gravity. The Higgs field is still non-minimally
coupled to gravity. However, the Higgs kinetic term is in a non-standard form.
In this way, we obtain (in the Einstein frame) a scalar field with a canonical kinetic term moving
in a potential having the desired exponential behaviour. The important
feature  we should emphasise here is that no approximation is needed.
\par
We start from the action
\bea
S_\text{J}= 
\int d^4x\sqrt{- {g}}\left[- \frac{1}{2}
\left(M^2+ {\xi}\,h^2  \right) R  +\frac{1}{2}
\,\frac{\xi\,h^2}{\left(M^2+ {\xi}\,h^2\right) }\,
\dd_\mu h\dd^\mu h
-\frac{\lambda}{4}\left(h^2-v^2\right)^2 \right]
 \,\,\,.
\label{my-model}
\eea
This means that the gauged kinetic term for the Higgs fields is modified as 
\bea
\frac{\xi\,\left(H^\dagger H\right)}{\left[M^2+ {2\xi}\,\left(H^\dagger H\right) \right] } 
\left(D_\mu H\right)^\dagger\left(D^\mu H\right)\,\,\,\,.
\label{DH-DH}
\eea
Here $D_\mu$ is the usual $SU\left(2\right)\times U\left(1\right)$ gauge covariant derivative.
It is important to notice that in the limit  $h^2>> \frac{M^2}{\xi}$, the two 
models (\ref{my-model}) and (\ref{B-S-0}) coincide. 
\par
We reach the Einstein frame by rescaling the metric as 
\bea
g_{\mu\nu}  \longrightarrow  e^{\omega\varphi} \,{g}_{\mu\nu}\,\,\,\,\,\,\,\,\,\,\,\,\text{with}\,\,\,\,\,\,\,\,\,\,\,\,
 \left(M^2+  {\xi}{{h}^2}  \right)e^{\omega\varphi} &=& {M^2_\text{p}}
\,\,\,\,.
\eea
Keeping as our variable the scalar field $\varphi$, we arrive at the action (up to a total derivative)
\bea
S_\text{E} &=& 
\int d^4x\sqrt{- {g}}\bigg\{
- \frac{M^2_\text{p}}{2}\, {R} 
+\frac{1}{2} {g}^{\mu\nu}\dd_\mu \varphi \dd_\nu \varphi
- \frac{\lambda}{4}\left(\frac{M^2_\text{p}}{\xi}\right)^2\bigg(1-  e^{\omega\varphi} \bigg)^2 
\bigg\}
 \,\,\,.
\label{my-model-1}
\eea
The constant $\omega$ is fixed by the relation 
\bea
\omega =- \, \frac{2}{M_\text{p}}\sqrt{\frac{\xi}{\left(1+6\xi\right)}} \simeq
-\, \sqrt{\frac{2}{3}}\,\frac{1} {M_\text{p}} \,\,\,(\text{for}  \,\,\, \,\,\,\xi>>1)
\,\,\,.
\label{omega}
\eea
Again, the relation ${M_\text{p} }^2=M^2+\xi\,v^2$ has been used.
The potential of the scalar field $\varphi$ varies slowly (the slow-roll condition)  when $ e^{\omega\varphi} << 1$. 
That is, inflation takes place for a large Higgs field $h$ 
as deduced from $\left(M^2+  {\xi}{{h}^2}  \right)e^{\omega\varphi} = {M^2_\text{p}}$.
\par
The theory defined by the action (\ref{my-model-1}) has been obtained without any approximation. On the contrary,
for the  inflationary model defined by  (\ref{B-S-apprx}) the approximations 
$\frac{M^2}{M^2_\text{p}}\, e^{\omega\varphi} << 1$ and $\xi>>1$ are assumed \cite{bezrukov1,bezrukov2}. 

\par
As usual, the first and second slow-roll parameters are defined by \cite{liddle, tsujikawa,senatore,mughal,julien}
\bea
\epsilon=  \frac{M_\text{p}^2}{2}\,\left( \frac{V^\prime}{V} \right)^2
\,\,\,\,\,\,\,\,\,\,\,,\,\,\,\,\,\,\,\,\,\,\,
\eta = {M_\text{p}^2} \,\frac{V^{\prime\prime}}{V} \,\,\,\,,
\eea
where a prime stands for the derivative with respect to $\varphi$ of the potential 
\be
V\left(\varphi\right)=\frac{\lambda}{4}\left(\frac{M^2_\text{p}}{\xi}\right)^2\bigg(1-  e^{\omega\varphi} \bigg)^2\,\,\,\,.
\ee
The inflationary observables are given by
\be
A_s= \frac{1}{24\pi^2 M_\text{p}^4}\,\frac{V}{\epsilon}\,\,\,\,\,,\,\,\,\,\,\,\,
n_s=1 + 2\eta - 6\epsilon \,\,\,\,\,,\,\,\,\,\,\,\,
r=16\epsilon \,\,\,\,.
\ee
Explicitly, we have
\bea
A_s &=& \frac{1} {192\pi^2 M_\text{p}^2\omega^2}\,\frac{\lambda}{\xi^2}\,
\frac{\left(1-  e^{\omega\varphi} \right)^4}  {e^{2\omega\varphi} }\,\,\,,
\nonumber \\
n_s &=& 1 - {4 M_\text{p}^2\omega^2}\,\frac{e^{\omega\varphi}\left(1+e^{\omega\varphi} \right) }{\left(1-  e^{\omega\varphi} \right)^2 }\,\,\,,
\nonumber \\
r  &=&  {32 M_\text{p}^2\omega^2}\,\frac{e^{2\omega\varphi} }{\left(1-  e^{\omega\varphi} \right)^2 }
\,\,\,.
\label{cosmo-parameters}
\eea
The quantities $A_s$, $r$ and $n_s$  are evaluated at $\varphi=\varphi_*$ defined
through 
\bea
N_*=  \frac{1} { M_\text{p}}\,\int_{\varphi_\text{end}}^{\varphi_*}\frac{d\varphi}{\sqrt{2\epsilon}}
= \frac{1} {2 M_\text{p}^2\omega^2}\bigg[\left(e^{-\omega\varphi_*}+\omega\varphi_*\right)
-\left(e^{-\omega\varphi_{\text{end}}}+\omega\varphi_{\text{end}}\right)\bigg]
\,\,\,.
\label{e-fold}
\eea
Here $N_*$ is the number of e-folds
and  ${\varphi_\text{end}}$ is the value of the field at which inflation ends. This is determined by
the condition $\epsilon\left( {\varphi_\text{end}}\right)=1$ and is found to be given by
\be
{\varphi_\text{end}}=-\frac{1}{\omega}\ln\left(1-\sqrt{2} { M_\text{p}}\omega\right)
\,\,\,.
\ee
If one uses the approximation ${ M_\text{p}}\omega\simeq-\, \sqrt{\frac{2}{3}}$, as given in (\ref{omega}), then
${\varphi_\text{end}}\simeq 0.9402{ M_\text{p}}$.
\par
The computation of ${\varphi_*}$ follows from   (\ref{e-fold}) and we found
\bea
 {\omega\varphi_*} &=& W\left(-e^{-c}\right) +c \,\,\,,
\nonumber \\
c &\equiv& {2 M_\text{p}^2\omega^2}\,N_* + \left(e^{-\omega\varphi_{\text{end}}}+\omega\varphi_{\text{end}}\right)
\simeq \frac{4}{3}N_* + 1.3870
 \,\,\,.
\eea
Here $W$ is the Lambert functions satisfying $W(x)e^{W(x)}=x$. The constant $c\simeq 81.3870$  for $N_*=60$.
\par
Since in our case the constant $c$ is real and we have $-e^{-1} < -e^{-c} <0$ then $W_{-1}\left(-e^{-c}\right)$ could be either
the principal branch $W_0$ or the branch  $W_{-1}$. Using the defining relation of 
$W$ one gets 
\bea
e^{\omega\varphi_*} &=& -\frac{1}{W\left(-e^{-c}\right)}\,\,\,.
\eea
This is what one needs for the determination of  
the parameters $\epsilon$, $\eta$ and $A_s$ given in (\ref{cosmo-parameters}). See also \cite{rubio} for a semilar treatment.

\par
As $-e^{-c}$ is of the order of $-10^{-36}$, the two branches of the Lambert function  $W\left(-e^{-c}\right)$
have, around zero,  the expansions \cite{gradshteyn,abramowitz,nist}
\bea
 W_0\left(-e^{-c}\right) &=& -e^{-c}-\left(-e^{-c}\right)^2+\dots\,\,\,\,,
\nonumber \\
 W_{-1}\left(-e^{-c}\right) &=& {-c}-\ln\left(c\right)+\dots\,\,\,\,.
\eea
Choosing the branch $W_0$ leads to an extremely big value for $e^{\omega\varphi_*}$. Therefore, we
take $e^{\omega\varphi_*}= -1/W_{-1}\left(-e^{-c}\right)$ and to first order we have
$e^{\omega\varphi_*}\simeq 1/c$.

\par
Substituting $e^{\omega\varphi}$ by approximately $1/c$ in (\ref{cosmo-parameters}) and assuming that $c>>1$,  
we find that
\bea
A_s & \simeq & \frac{1} {192\pi^2 M_\text{p}^2\omega^2}\,\frac{\lambda}{\xi^2}\,c^2
\simeq  \frac{1} {72\pi^2 }\,\frac{\lambda}{\xi^2}\,N_*^ 2\,\,\,\,.
\eea
In the last expression we have taken $c\simeq \frac{4}{3}N_*$. The experimental value for $A_s$ is
around \cite{planck-2018}
\bea
A_s \simeq 10^{-10}\,e^{3.094} \,\,\,\,.
\eea
This fixes the value  of  the parameter $\xi$ as
\be
\xi\simeq 600\, c\,\sqrt{\lambda} \simeq 800 N_*\sqrt{\lambda} \,\,\,.
\ee
Taking $N_*\simeq 60$ \cite{planck-2018} and assuming that the Higgs self-coupling is $\lambda =0.15$, we get the value
\bea
\xi\simeq 1.86\times 10^{4 } \,\,\,.
\eea
In the same way and for $N_*=60$, the quantities $n_s$ and $r$ are approximately given by
\bea
n_s &\simeq& 1-\frac{2}{N_*}\simeq 0.9667 \,\,\,\,,
\nonumber \\
r &\simeq& \frac{12}{N_*^2}\simeq 0.0033 \,\,\,\,.
\eea
These nicely agree with the measured values \cite{planck-2018}.

\par
Before leaving this section, let us explore how does the field  $\varphi$
couple to the other fields of the Standard Model.
In the unitary gauge and according to (\ref{DH-DH}), a generic coupling between the Higgs field and any  gauge field is of
the form\footnote{When the Standard Model is coupled to gravity, all the fermionic kinetic terms as well as the
gauge kinetic terms are invariant under the local rescalings
$g_{\mu\nu} \, \longrightarrow\, e^{\omega\varphi} \,{g}_{\mu\nu}$, $A_\mu \, \longrightarrow\,A_{\mu}$, 
 $ \psi_{L,R} \, \longrightarrow\, e^{-\frac{3}{4}\omega\varphi}\,\psi_{L,R}$.  
The spacetime dependent Dirac $\gamma^\mu$ matrices are also rescaled as $ \gamma^\mu \, \longrightarrow\, e^{-\frac{1}{2}\omega\varphi} \,\gamma^\mu$.}
\bea
\sqrt{-g} \,\frac{\xi\,h^2}{\left(M^2+ {\xi}\,h^2\right)}\, h^2  A_\mu A_\nu g^{\mu\nu}\,\,\,\,.
\eea
Under the rescalings $g_{\mu\nu} \, \longrightarrow\, e^{\omega\varphi} \,{g}_{\mu\nu}$, 
$A_\mu \, \longrightarrow\,A_{\mu}$ with 
$\left(M^2+  {\xi}{{h}^2}  \right)e^{\omega\varphi} = {M^2_\text{p}}$, this coupling term
becomes
\bea
\sqrt{-g} \,\frac{\xi\,h^2}{\left(M^2+ {\xi}\,h^2\right)}\, h^2  A_\mu A_\nu g^{\mu\nu}
\, \longrightarrow\,
\frac{M^2_\text{p}}{\xi}\left(1-\frac{M^2}{M^2_\text{p}}\,e^{\omega\varphi}\right)^2
 \sqrt{-{g}} \,   A_\mu A_\nu {g}^{\mu\nu}\,\,\,\,.
\eea
We see that in the inflationary phase when $\frac{M^2}{M^2_\text{p}}\,e^{\omega\varphi} << 1$  a decoupling between the fields
$\varphi$ and $A_\mu$ takes place. 
\par
The  Yukawa terms have not been modified in this model. A generic 
mass generating Yukawa interaction is still given by
\bea
\sqrt{-g}\,h\left(\bar{\psi}_L\,\chi_R +  \bar{\chi}_R \psi_L\right)\,\,\,\,,
\eea
where $\psi_L$ is a left-handed fermion   and $\chi_R$  is a right-handed fermion.
Under the rescalings  $g_{\mu\nu} \, \longrightarrow\, e^{\omega\varphi} \,{g}_{\mu\nu}$, 
 $(\psi_L,\chi_R) \, \longrightarrow\, e^{-\frac{3}{4}\omega\varphi}(\psi_L,\chi_R)$ together with 
$\left(M^2+  {\xi}{{h}^2}  \right)e^{\omega\varphi} = {M^2_\text{p}}$, this fermionic coupling term
transforms into
\bea
\sqrt{-g}\,h\left(\bar{\psi}_L\,\chi_R +  \bar{\chi}_R \psi_L\right)
\, \longrightarrow\,
\frac{M_\text{p}}{\sqrt{\xi}}\left(1-\frac{M^2}{M^2_\text{p}}\,e^{\omega\varphi}\right)^{\frac{1}{2}}
 \sqrt{-\hat{g}}\left(\bar{\psi}_L\,\chi_R +  \bar{\chi}_R \psi_L\right)\,\,\,\,.
\eea
The interaction between the field $\varphi$ and the fermions is highly suppressed during inflation.

\section{Conclusions}

There is a freedom when it comes to coupling the Standard Model of particle physics to gravity.
Since the inclusion of gravity leads to non-renormalisability of the theory, the Standard Model
coupled to gravity might therefore be viewed as an effective theory. In this note we have exploited
this freedom and proposed a viable effective theory for describing the inflationary regime of the early
Universe. We have modified the model, advocated in ref.\cite{bezrukov1},  where the Higgs field is non-minimaly coupled to  gravity.
The guiding principle to this modification is that it results in $i)$ a canonical kinetic term for the inflaton
field and  $ii)$ a simple 
slow-roll potential for this field. We have indeed succeeded in meeting these two requirements without any approximation.
This is achieved by modifying the Higgs kinetic term. Our model is therefore a completion of the effective
theory proposed in \cite{bezrukov1}. It totally agrees with  its  cosmological findings.
\par
Our study could be extended to the Higgs-dilaton models of gravity \cite{shapo-1,shapo-2,shapo-3,shapo-4}. {}For this, we consider
the action
\bea
S_\text{J} &=& 
\int d^4x\sqrt{-{g}}\left[ - \frac{1}{2}\left(
\tau\,\sigma^2 +{\xi}\,h^2\right)R +\frac{A}{2}\,\dd_\mu \sigma\dd^\mu \sigma  +\frac{B}{2}\,\dd_\mu h\dd^\mu h
+ {C}\,\,\dd_\mu \sigma\dd^\mu h \right.
\nonumber \\
&-& \left. \frac{\lambda}{4}\left(h^2-\alpha\sigma^2\right)^2 -\beta\sigma^4 \right]
 \,\,\,.
\label{D-H}
\eea
The quantities $A$, $B$ and $C$ are functions of the two scalar fields $\sigma$ and $h$. 
The Higgs-dilaton model correspond to $A=B=1$ and $C=0$ \cite{shapo-1,shapo-2,shapo-3,shapo-4}. 
\par
As usual, we rescal the metric as 
\bea
g_{\mu\nu}  \longrightarrow  e^{\omega\varphi} \,{g}_{\mu\nu}\,\,\,\,\,\,\,\,\,\,\,\,\text{with}\,\,\,\,\,\,\,\,\,\,\,\,
 \left(\tau\,\sigma^2+  {\xi}{{h}^2}  \right)e^{\omega\varphi} &=& {M^2_\text{p}}
\,\,\,\,.
\eea
The aim now is to find the functions $A$, $B$ and $C$ such that two of the scalar fields have canonical kinetic
terms and their corresponding  potential is simple. As the three fields    $\sigma$, $h$ and $\varphi$ are related by
the last relation, we choose to keep the two fields $\sigma$ and $\varphi$. The functions   $A$, $B$ and $C$
that fulfill these requirements are
\bea
A &=& \frac{\tau\,\sigma^2 +{\xi}\,h^2 }{{M^2_\text{p}} }
+ \frac{4\tau^2\left(1-\frac{2}{3} \omega^2{M^2_\text{p}}\right)}{\omega^2{M^2_\text{p}} }
 \frac{\sigma^2}{\tau\,\sigma^2 +{\xi}\,h^2 }\,\,\,\,,
\nonumber \\
B &=& 
 \frac{4\xi^2\left(1-\frac{2}{3} \omega^2{M^2_\text{p}}\right)}{\omega^2{M^2_\text{p}} }
 \frac{h^2}{\tau\,\sigma^2 +{\xi}\,h^2 }\,\,\,\,,
\nonumber \\
C &=& 
 \frac{4\tau\,\xi\left(1-\frac{2}{3} \omega^2{M^2_\text{p}}\right)}{\omega^2{M^2_\text{p}} }
 \frac{\sigma\,h}{\tau\,\sigma^2 +{\xi}\,h^2 }\,\,\,\,.
\eea
With these expression for  $A$, $B$ and $C$, the above rescaling of the metric
yields the action
\bea
S_\text{E} &=& 
\int d^4x\sqrt{-{g}}\left[ - \frac{{M^2_\text{p}} }{2}\,R 
+\frac{1}{2}\,\dd_\mu \sigma\dd^\mu \sigma  +\frac{1}{2}\,\dd_\mu \varphi\dd^\mu \varphi 
\right.
\nonumber \\
 &-& \left. \frac{\lambda}{4}\left(\frac{{M^2_\text{p}} }{\xi} \right)^2\left[
1 - \frac{\xi}{{M^2_\text{p}} }\left(\alpha + \frac{\tau}{\xi}\right)\sigma^2\, e^{\omega\varphi} \right]^2
-\beta\,\left(\sigma^2\, e^{\omega\varphi}  \right)^2 \right]
 \,\,\,.
\label{D-H-1}
\eea
Although the first term in the expression of $A$ breaks the global scale invariance, the resulting action
in the Einstein frame is simple. It is appealing to investigate the cosmological consequences 
of this theory along the lines of ref.\cite{rubio-1}. We will report on this project elsewhere.
It is also worth treating the case  $A=B=1$ and $C={\text{constant}}$,  which preserves the global 
scale invariance.


\begin{thebibliography}{99}


\bibitem{liddle}

Andrew R. Liddle, {\it An introduction to cosmological inflation}, arXiv:astro-ph/9901124.
	

\bibitem{tsujikawa}

Shinji Tsujikawa, {\it Introductory review of cosmic inflation}, (2003), arXiv:hep-ph/0304257.

\bibitem{senatore}

Leonardo Senatore,  {\it Lectures on Inflation}, arXiv:1609.00716 [hep-th].

\bibitem{mughal}

Muhammad Zahid Mughal, Iftikhar Ahmad and Juan Luis Garc\'{\i}a Guirao,  {\it Relativistic Cosmology with an Introduction to Inflation},
Universe 2021, {\bf 7}, 276. https://doi.org/10.3390/universe7080276.


\bibitem{julien}
Julien Lesgourgues, {\it Inflationary cosmology}, (2006), 
\newline
https://lesgourg.github.io/courses.html.

\bibitem{steinwachs}

Christian F. Steinwachs, {\it Higgs field in cosmology}, arXiv:1909.10528 [hep-ph].

\bibitem{rubio}
Javier Rubio, {\it Higgs inflation}, Front. Astron. Space Sci.  {\bf 5:50} (2019), arXiv:1807.02376 [hep-ph].


\bibitem{horn}
Bart Horn, 
{\it The Higgs Field and Early Universe Cosmology: A (Brief) Review}, Physics 2020, {\bf 2(3)}, 503-520,
arXiv:2007.10377 [hep-ph].

\bibitem{kaiser}
David Kaiser, {\it Primordial Spectral Indices from Generalized Einstein Theories}, Phys. Rev. {\bf D 52} (1995) 4295-4306,	arXiv:astro-ph/9408044.


\bibitem{bezrukov1}
Fedor L. Bezrukov and Mikhail Shaposhnikov, {\it The Standard Model Higgs boson as the inflaton}, 
Phys. Lett. {\bf B 659} (2008) 703,  arXiv:0710.3755 [hep-th].

\bibitem{bezrukov2}

Fedor Bezrukov, {\it The Higgs field as an inflaton}, Class. Quantum Grav. {\bf 30} (2013) 214001, arXiv:1307.0708 [hep-ph].

\bibitem{starobinsky}
Alexei A. Starobinsky,
{\it A New Type of Isotropic Cosmological Models Without Singularity}, Phys. Lett. {\bf B 91} (1980) 99.

\bibitem{kehagias}
Alex Kehagias, Azadeh Moradinezhad Dizgaha and Antonio Riottoa,
{\it Comments on the Starobinsky Model of Inflation and its Descendants},
Phys. Rev. {\bf D 89} (2014) 043527, arXiv:1312.1155 [hep-th].

\bibitem{whitt}
Brian Whitt, {\it Fourth Order Gravity as General Relativity Plus Matter}, Phys. Lett. {\bf B 145} (1984) 176-178.



\bibitem{gradshteyn}
{\it Table of Integrals, Series, and Products}, I. S. Gradshteyn and I. M. Ryzhik,
(Alan Jeffrey, Editor), Academic Press, Fifth Edition (1994). 


\bibitem{abramowitz}
{\it Handbook of Mathematical Functions}, Edited by  M. Abramowitz and I. A. Stegun,
Dover Publications, New York, Ninth Printing (1970).


\bibitem{nist}
{\it NIST Handbook of Mathematical Functions}, Edited by F. W. J. Olver, D. W. Lozier, R. F. Boisvert and C. W. Clark,
Cambridge University Press,  New York, First Published (2010).

\bibitem{planck-2018}

Planck Collaboration: Y. Akrami et al., {\it Planck 2018 results. X. Constraints on inflation}, Astronomy $\&$ Astrophysics 641, {\bf A10} (2020), arXiv:1807.06211 [astro-ph.CO].



\bibitem{shapo-1}

Mikhail Shaposhnikov, Daniel Zenh\"ausern,
{\it Scale invariance, unimodular gravity and dark energy}, Phys. Lett. {\bf B 671 } (2009) 187-192,
arXiv:0809.3395 [hep-th].


\bibitem{shapo-2}

Juan Garc\'{\i}a-Bellido, Javier Rubio, Mikhail Shaposhnikov and  Daniel Zenh\"ausern, {\it Higgs-Dilaton Cosmology: From the Early to the Late Universe}, 
Phys. Rev. {\bf D 84} (2011) 123504, arXiv:1107.2163 [hep-ph].


\bibitem{shapo-3}

Diego Blas, Mikhail Shaposhnikov and Daniel Zenh\"ausern, {\it Scale-invariant alternatives to general relativity}, 
Phys. Rev. {\bf D 84}  (2011) 044001, arXiv:1104.1392 [hep-th].

\bibitem{shapo-4}

Fedor Bezrukov, Georgios K. Karananas, Javier Rubio and Mikhail Shaposhnikov, {\it Higgs-Dilaton Cosmology: an effective field theory approach}, 
Phys. Rev. {\bf D 87} (2013) 096001, arXiv:1212.4148 [hep-ph].



\bibitem{rubio-1}

Santiago Casas, Martin Pauly and  Javier Rubio,
{\it Higgs-Dilaton Cosmology: An inflation - dark energy connection and forecasts for future galaxy surveys},
Phys. Rev. {\bf D 97} (2018) 043520, arXiv:1712.04956 [astro-ph.CO].







\end{thebibliography}
\end{document}